\pdfoutput=1
\documentclass[aps,prl,twocolumn, titlepage,showpacs]{revtex4}

\usepackage{graphicx}
\usepackage{color}
\usepackage{dcolumn}

\usepackage{bm}

\begin{document}

\title{Energy Transport in a Shear Flow of Particles in a 2D Dusty Plasma}

\author{Yan Feng}
\email{yan-feng@uiowa.edu}
\author{J. Goree}
\author{Bin Liu}
\affiliation{Department of Physics and Astronomy, The University
of Iowa, Iowa City, Iowa 52242, USA}

\date{\today}

\begin{abstract}

A shear flow of particles in a laser-driven two-dimensional (2D) dusty plasma are observed in a further study of viscous heating and thermal conduction. Video imaging and particle tracking yields particle velocity data, which we convert into continuum data, presented as three spatial profiles: mean particle velocity (i.e., flow velocity), mean-square particle velocity, and mean-square fluctuations of particle velocity. These profiles and their derivatives allow a spatially-resolved determination of each term in the energy and momentum continuity equations, which we use for two purposes. First, by balancing these terms so that their sum  (i.e., residual) is minimized while varying viscosity $\eta$ and thermal conductivity $\kappa$ as free parameters, we simultaneously obtain values for $\eta$ and $\kappa$ in the same experiment. Second, by comparing the viscous heating and thermal conduction terms, we obtain a spatially-resolved characterization of the viscous heating.

\end{abstract}

\narrowtext

\maketitle

\section {I.~INTRODUCTION}
Flows of most liquid substances are usually studied by modeling the liquid as a continuum, but there are some substances that allow the study of flows at the kinetic level, i.e., at the level of the individual constituent particles. As examples, we can mention chute flows in granular materials~\cite{Pouliquen:96} and capillary flows in colloids~\cite{Isa:07}. The solid particles in these soft materials are large enough that their motion can be tracked by video microscopy, allowing experimenters to record their positions and velocities. Like granular materials and colloids, dusty plasmas also allow direct observation of individual particle motion.

Dusty plasma~\cite{Melzer:08, Morfill:09, Piel:10, Shukla:02, Bonitz:10} is a four-component mixture consisting of micron-size particles of solid matter, neutral gas atoms, free electrons, and free positive ions. These particles of solid matter, which are referred to as ``dust particles,'' gain a large negative charge $Q$, which is about {$- 10^4$} elementary charges under typical laboratory conditions. The motion of the dust particles is dominated by electric forces, corresponding to the local electric field ${\bf E} = {\bf E}_{\rm conf} + {\bf E}_d$, where ${\bf E}_{\rm conf}$ is due to confining potentials and ${\bf E}_d$ is due to Coulomb collisions with other dust particles.

Due to their high charges, Coulomb collisions amongst dust particles have a dominant effect. The interaction force $Q{\bf E}_d$ amongst dust particles is so strong that the dust particles do not move easily past one another, but instead self-organize and form a structure that is analogous to that of atoms in a solid or liquid~\cite{Chu:94, Thomas:96, Melzer:96, Samsonov:04, Feng:08, Sheridan:08, Feng:10_2, Hartmann:10}. In other words, the collection of dust particles is said to be a strongly-coupled plasma~\cite{Ichimaru:82}. In a strongly-coupled plasma, the pressure $p$ is due mainly to interparticle electric forces, with only a small contribution from thermal motion~\cite{Flanagan:10}.

Even when it has a solid structure, a collection of dust particles is still very soft, as characterized by a sound speed on the order of 1~cm/s~\cite{Nunomura:02, Feng:10}. As a result, a dusty plasma in a solid phase is very easily deformed by small disturbances, and it can be made to flow. Flows can be generated, for example, by applying shear using a laser beam that exerts a spatially-localized radiation force~\cite{Nosenko:04, Feng:10, Liu:03, Homann:98, Juan:01, Wolter:05, Nunomura:05, Vaulina:08, Fink:11}. In such an experiment, the Reynolds number is usually very low, typically $R_e \approx 10$, indicating that the flow is laminar~\cite{Nosenko:04}.

This paper provides further analysis and details of the experiment that was reported in~\cite{Feng:12}. We now list some of the major points of these two papers, to indicate how they are related and how they differ. In this paper, we present: (1) a detailed treatment of the continuity equations for both momentum and energy, (2) our method of simultaneously determining two transport coefficients (viscosity and thermal conductivity), (3) values of these two coefficients, and (4) spatially-resolved profiles of the terms of the energy equation, including the terms for viscous heating and thermal conduction, as determined by experimental measurements. In~\cite{Feng:12}, we reported: (1) a discovery of peaks in a spatially-resolved measurement of kinetic temperature, (2) a demonstration that these peaks are due to viscous heating in a region of a highly sheared flow velocity, and (3) a quantification of the role of viscous heating, in competition with thermal conduction, by reporting a dimensionless number of fluid mechanics called the Brinkman number~\cite{Huba:94} which we found to have an unusually large value due to the extreme properties of dusty plasma as compared to other substances. The values of viscosity and thermal conduction found in this paper are used as inputs for the calculations of dimensionless numbers in~\cite{Feng:12}. The identification of viscous heating as the cause of the temperature peaks reported in~\cite{Feng:12} is supported by the spatially-resolved measurements reported here.

In the experiment, the dust particles are electrically levitated and confined by the electric field in a sheath above a horizontal lower electrode in a radio-frequency (rf) plasma, forming a single layer of dust particles, Fig.~1(a). The dust particles can move easily within their layer, but very little in the perpendicular direction, so that their motion is mainly two dimensional (2D). They interact with each other through a shielded Coulomb (Yukawa) potential, due to the screening provided by the surrounding free electrons and ions~\cite{Konopka:00}. As the dust particles move, they also experience frictional drag since individual dust particles in general move at velocities different from those of the molecules of the neutral gas. This friction can be modeled as Epstein drag~\cite{Liu:03}, and characterized by the gas damping rate $\nu_{\rm gas}$.

Using experimental measurements of their positions and velocities, the dust particles can of course be described in a {\it particle} paradigm. They can also be described by a {\it continuum} paradigm by averaging the particle data on a spatial grid. In transport theory, momentum and energy transport equations are expressed in a continuum paradigm, while transport coefficients such as viscosity and thermal conductivity are derived using the particle paradigm because these transport coefficients are due to collisions amongst individual particles. In our experiment, we average the data for particles, such as velocities, to obtain the spatial profiles for the continuum quantities, such as flow velocity. In the continuum paradigm, a substance obeys continuity equations that express the conservation of mass, momentum, and energy. These continuity equations, which are also known as Navier-Stokes equations, characterize the transport of mass, momentum, and energy. In a multi-phase or multi-component substance, these equations can be written separately for each component. The component of interest in this paper is dust.

In Sec.~II, we review the continuity equations for dusty plasmas. In Secs.~III and IV, we provide details of our experiment and data analysis method. We designed our experiment to have significant flow velocity and significant gradients in the flow velocity, i.e., velocity shear. In Sec.~V, we simplify the continuity equations using the spatial symmetries and steady conditions of the experiment, and including the effects of external forces. We will use our experimental data as inputs in these simplified continuity equations in Secs.~VI and VII.

\section {II.~Continuity equations for dusty plasmas}
We now review the continuity equations for mass, momentum, and energy for dusty plasmas. We will then discuss the significance of some of the terms for our experiments.

The equation of mass continuity, i.e., conservation of mass, is
\begin{equation}\label{mass}
{\frac{\partial {\rho}}{\partial t} + \nabla \cdot (\rho {\bf v}) = 0.}
\end{equation}
In this paper, the mass density $\rho$ and the fluid velocity ${\bf v}$ describe the dust continuum.

The momentum equation~\cite{Batchelor:67, Landau:87} is
\begin{eqnarray}\label{momentum}
\frac{\partial {\bf v}}{\partial t} + {\bf v} \cdot \nabla {\bf v} = \frac{\rho_c {\bf E_{\rm conf}}}{\rho} - \frac{\nabla p}{\rho} & + & \frac{\eta}{\rho} \nabla^2 {\bf v}
\nonumber\\
 + \left[ \frac{\zeta}{\rho} + \frac{\eta}{3\rho} \right]\nabla(\nabla \cdot {\bf v}) & + & {\bf f}_{\rm ext},
\end{eqnarray}
Here, $\rho_c$, $\eta$, and $\zeta$ are the charge density, shear viscosity, and bulk viscosity, respectively; and $\eta / \rho$ is called the kinematic viscosity. In this paper, these parameters describe the dust continuum. Equation~(\ref{momentum}) describes the force per unit mass, i.e., acceleration, for the continuum. The confining field ${\bf E_{\rm conf}}$ and pressure $p$ have been discussed in Sec.~I. The last term in Eq.~(\ref{momentum}) is due to the momentum contribution from forces such as gas friction, laser manipulation, ion drag, and any other forces that are external to the layer of dust particles, as discussed in Sec.~IV. The other terms on the right-hand-side of Eq.~(\ref{momentum}) correspond to viscous dissipation, which arises from Coulomb collisions amongst the charged dust particles. The viscous term $(\eta / \rho) \nabla^2 {\bf v}$ was studied in a previous dusty plasma experiment~\cite{Nosenko:04}.

For our experiment, we can consider Eq.~(\ref{momentum}) for two conditions, with and without the application of the external force ${\bf f}_{\rm ext}$. Without the external force, the fluid velocity ${\bf v}$ is zero so that Eq.~(\ref{momentum}) is reduced to $\rho_c {\bf E_{\rm conf}}/\rho -\nabla p/\rho = 0$, meaning that the confining electric force is in balance with the pressure (which mainly arises from interparticle electric forces). With the external force, the external confining force remains as it was without ${\bf f}_{\rm ext}$; moreover, the pressure $p$ will be affected only weakly because the density is unchanged (as we will show later) so that the interparticle electric forces that dominate the pressure will also be unchanged. Therefore, in using Eq.~(\ref{momentum}), we will assume that the first two terms on the right hand side always cancel everywhere.

The internal energy equation, as it is expressed commonly in fluid dynamics~\cite{Batchelor:67, Landau:87}, is
\begin{equation}\label{energy}
{T\left(\frac{\partial s}{\partial t} + {\bf v} \cdot \nabla s\right) = \Phi + \frac {\kappa}{\rho} \nabla^2 T + P_{\rm ext}.}
\end{equation}
Here, $s$ is the entropy per unit mass, $\kappa$ is the thermal conducitivity, and $T$ is the thermodynamic temperature of the dust continuum. The last term $P_{\rm ext}$ is due to the energy contribution from any external forces ${\bf f}_{\rm ext}$.

We assume that the continuity equations, Eqs.~(\ref{mass}-\ref{energy}), are valid for the dust particles separately from other components of dusty plasmas that occupy the same volume, such as neutral gas atoms. The coupling between the dust particles with other components is indicated in ${\bf f}_{\rm ext}$ and $P_{\rm ext}$, so that the momentum and energy of the dust particle motion is treated for the dust particles separately from other components. Other external forces, such as those due to laser manipulation, are also indicated in ${\bf f}_{\rm ext}$ and $P_{\rm ext}$.

The first term on the right-hand-side of Eq.~(\ref{energy}) is due to viscous heating. The viscous heating term $\Phi$ depends on the square of the shear, i.e., the square of the gradient of flow velocity. A general expression for $\Phi$ has many terms (cf. Eq.~(3.4.5) of Ref.~\cite{Batchelor:67} or Eq.~(49.5) of Ref.~\cite{Landau:87}), but it can be simplified for our experiment by taking advantage of symmetries, as explained in Sec.~IV.

The second term on the right-hand-side of Eq.~(\ref{energy}) is due to thermal conduction. It arises from a temperature gradient. Previous experiments with 2D dusty plasmas include a study of the thermal conduction term in Eq.~(\ref{energy})~\cite{Nosenko:08}.

In this paper, most of our attention will be devoted to the first two terms on the right-hand-side of Eq.~(\ref{energy}). Using our experimental data, we will compare the magnitudes of these terms. In~\cite{Feng:12}, we demonstrated that viscous heating is measurable and significant, when evaluated using only {\it global} measures like the Brinkman number. Here we further evaluate viscous heating by characterizing it {\it locally} using spatially-resolved profiles for the terms in Eq.~(\ref{energy}). We also develop a method to simultaneously obtain values of two transport coefficients of $\eta$ and $\kappa$.

\section {III.~Experiment}

Here we provide a more detailed explanation of the experiment than in~\cite{Feng:12}. An argon plasma was generated in a vacuum chamber at $15.5~{\rm mTorr}$ (or 2.07 Pa), powered by rf voltages at $13.56~{\rm MHz}$ and $214~{\rm V}$ peak-to-peak. We used the same chamber and electrodes as in~\cite{Feng:11}. The dust particles were $8.09~{\rm \mu m}$ diameter melamine-formaldehyde microspheres of mass $m_d = 4.18 \times 10^{-13}~{\rm kg}$. The dust particles settled in a single layer above the powered lower electrode. The layer of dust particles had a circular boundary with a diameter of $\approx\,52~{\rm mm}$ and contained $\approx\,10^4$ dust particles. As individual dust particles moved about within their plane, they experienced a frictional damping~\cite{Liu:03} with a rate $\nu_{\rm gas} = 2.7~{\rm s^{-1}}$ due to the surrounding argon gas.

The particles were illuminated by a $488$-nm argon laser beam that was dispersed to provide a thin horizontal sheet of light, Fig.~1(a). Using a cooled 14-bit digital camera (PCO model 1600) viewing from above, we recorded the motion of individual dust particles. This top-view camera imaged a central portion of the dust layer, as sketched in Fig.~1(b). The movie is available for viewing in the Supplemental Material of~\cite{Feng:12}. The portion of the camera's field of view that we will analyze was $23.5 \times 23.5~{\rm mm^2}$, and it contained $\approx 2500$ particles. We recorded $> 5000$ frames at a rate of 55 frame/s, with a resolution of $0.039~{\rm mm/pixel}$. Our choice of 55 frame/s was sufficient for accurate measurement of various dynamical quantities, including the kinetic temperature, although a slightly higher frame rate would have been optimal~\cite{Feng:11_2}. In addition to the top-view camera, we also operated a side-view camera to verify that there was no significant out-of-plane motion; this was due to a strong vertical confining electric field. Thus, we will analyze the particle motion data taking into consideration only the motion within a horizontal plane.

At first, the dust particles self-organized in their plane to form a 2D crystalline lattice. The particle spacing, as characterized by a Wigner-Seitz radius~\cite{Kalman:04}, was $a = 0.26~{\rm mm}$, corresponding to a lattice constant $b = 0.50~{\rm mm}$, an areal number density $n_0 = 4.7~{\rm mm^{-2}}$, and a mass density $\rho = n m_d = 1.97 \times 10^{-12}~{\rm kg/mm^{2}}$. Using the wave-spectra method for thermal motion of particles in the undisturbed lattice, we found the following parameters for the dust layer: $\omega_{pd} = 75~{\rm s^{-1}}$, $Q/e = - 9700$, and $a/\lambda_D = 0.5$, where $\omega_{pd}$ is the nominal 2D dusty plasma frequency~\cite{Kalman:04}, $Q$ is the particle charge, $e$ is the elementary charge, and $\lambda_D$ is the screening length of the Yukawa potential.

We used laser manipulation~\cite{Liu:03, Homann:98, Juan:01, Nosenko:04, Wolter:05, Nunomura:05, Vaulina:08, Fink:11} to generate stable flows in the same 2D dusty plasma layer. A pair of continuous-wave $532$-nm laser beams struck the layer at a $6^\circ$ downward angle. In this manner, we applied radiation forces that pushed particles in the $\pm x$ directions, as shown in Fig.~1. The power of each beam was $2.28~{\rm W}$ as measured inside the vacuum chamber. To generate a wider flow than in~\cite{Feng:10, Nosenko:04}, the laser beams were rastered in the both $x$ and $y$ directions in a Lissajous pattern as in~\cite{Nosenko:06}, with frequencies of $f_x = 123.607~{\rm Hz}$ and $f_y = 200~{\rm Hz}$. We chose these frequencies to be $\gg \omega_{pd}/2\pi$ to avoid exciting coherent waves. The rastered laser beams filled a rectangular region, which crossed the entire dust layer and beyond, as sketched in Fig.~1(b). This laser manipulation scheme resulted in a circulating flow pattern with three stable vortices, as sketched in Fig.~1(b). We designed the experiment so that in the analyzed region the flow is straight in the $\pm x$ directions, without any significant curvature. Curvature in the flow, which is necessary for the flow pattern to close on itself as sketched in Fig.~1(b), was limited in our experiment to the extremities of the dust layer, where it would not affect our observations.

\begin{figure}[htb]
\centering
\includegraphics{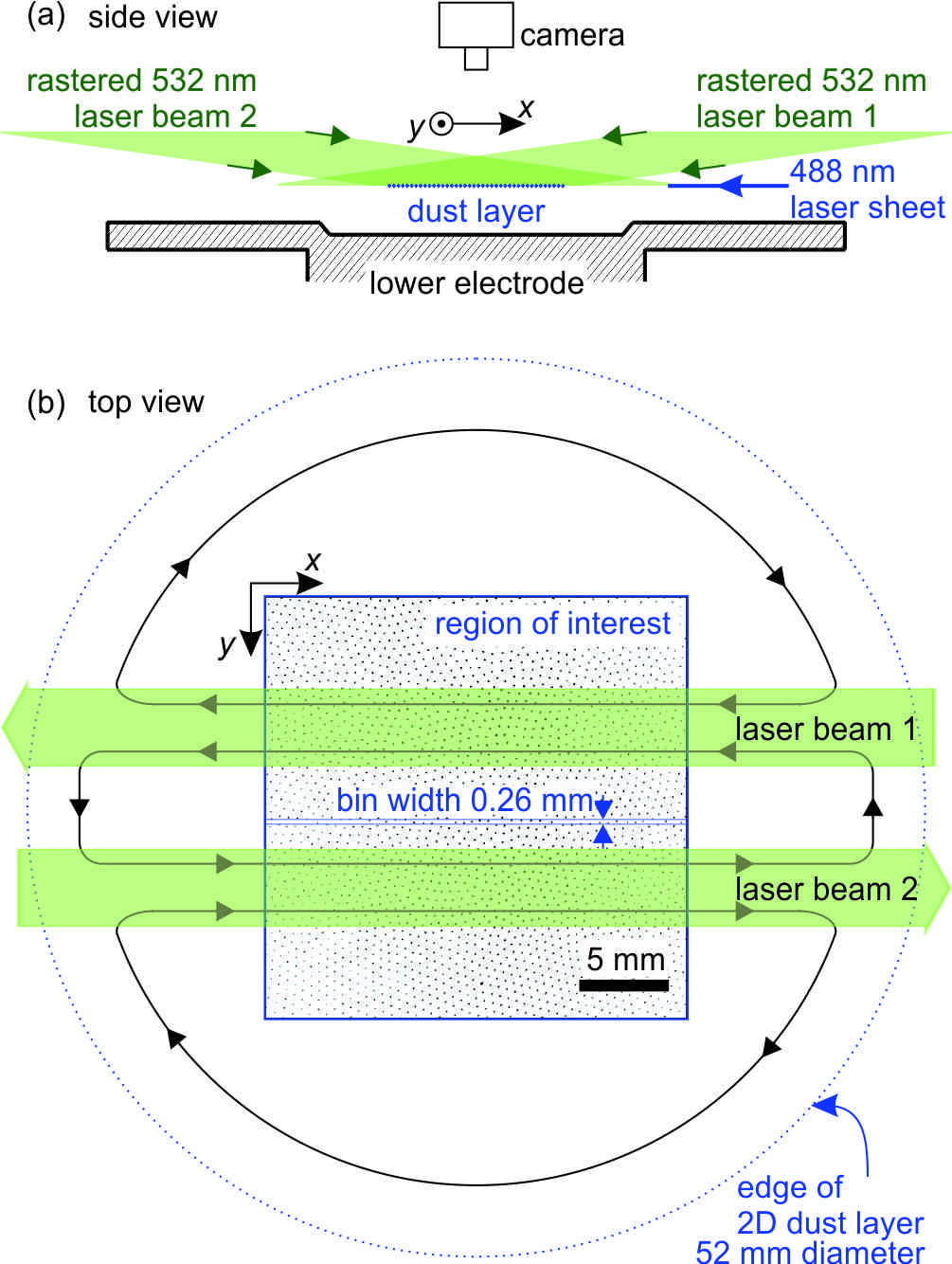}
\caption{\label{etasketch} (Color online). (a) Side-view sketch of the apparatus, not to scale. A single layer of dust particles of charge $Q$ and mass $m_d$ are levitated against gravity by a vertical dc electric field. There is also a weaker radial dc electric field $E_{\rm conf}$ which prevents the dust particles from escaping in the horizontal direction. Further details of the chamber are shown in~\cite{Feng:11}. The two laser beams are rastered in the $y$ direction so that they have a finite expanse, and they are offset in the $y$ direction as shown in (b). (b) Top-view sketch of laser-driven flows in the 2D dusty plasma. In the region of interest, the flow is straight, with curvature of the flow limited to the extremities of the dust layer. A video image of the dust particles within the region of interest is also shown. The region of interest is divided into 89 bins of width $0.26~{\rm mm}$ so that particle data can be converted to continuum data.}
\end{figure}

\section {IV.~Particle and Continuum Paradigms}

We start our data analysis by analyzing data for individual particles, i.e., by working in the {\it particle} paradigm. Using image analysis software~\cite{ImageJ}, with a method optimized as in~\cite{Feng:07} to minimize measurement error, we identify individual particles in each video image and calculate their $x-y$ coordinates. We then track a dust particle between two consecutive frames and calculate its velocity ${\bf v}_d$ as the difference in its position divided by the time interval between frames~\cite{Feng:11_2}. Now having the position and velocity of all the particles in the analyzed region, we can study motion at the particle level. For example, in Sec.~VI we will use data for the individual particles to calculate the rate of their energy dissipation due to their frictional drag with the neutral gas, $P_{\rm ext}$.

We next convert our data for individual particles to continuum data, i.e., we change from the {\it particle} paradigm to the {\it continuum} paradigm. This is done by averaging particle data within spatial regions of finite area, which we call bins. There are 89 bins, which are all long narrow rectangles aligned in the $y$ direction, as shown in Fig.~1(b). Each bin contains $\approx 30$ dust particles. We choose the shape of these bins to exploit the symmetry of the experiment, which in the analyzed region has an ignorable coordinate, $x$. The width of each bin is the same as the Wigner-Seitz radius, $a$. To reduce the effect of particle discreteness as a particle crosses the boundaries between bins, we use the cloud-in-cell weighting method~\cite{Birdsall:91, Feng:10} which has the effect of smoothing data so that a particle contributes its mass, momentum and energy mostly to the bin where it is currently located and to a lesser extent to the next nearest bin. The data are binned in this way regardless of their $x$ positions, since $x$ is treated as an ignorable coordinate. We also time-average these binned data, exploiting the steady conditions of the experiment. This procedure (binning, cloud-in-cell-weighting, and time averaging) yields our continuum quantities, such as the flow velocity $\overline{\bf v}(y)$. It also yields a kinetic temperature
\begin{equation}\label{KT}
{k_B T_{\rm kin} = \frac{1}{2} m_d \overline{|{\bf v}_d - \overline{\bf v}|^2},}
\end{equation}
which is calculated from the individual particle velocities; this kinetic temperature is not necessarily identical to the thermodynamic temperature $T$. Here, $k_B$ is the Boltzmann constant.

We assume that it is valid to use a continuum model when gradients are concentrated in a region as small as a few particle spacings. In fact, it has been shown experimentally that the momentum equation~\cite{Nosenko:04} and the energy equation~\cite{Nosenko:08} remain useful in 2D dusty plasma experiments with gradients that are as strong as in our experiment.

The notation we use in this paper distinguishes velocities and other quantities according to whether they correspond to individual particles or continuum quantities. Parameters for individual dust particles are denoted by a subscript $d$, for example ${\bf v}_d$ for the velocity of an individual dust particle. Continuum quantities in a theoretical expression are indicated without any special notation, for example ${\bf v}$ for the hydrodynamic velocities in Eqs.~(\ref{mass}) and (\ref{momentum}). Finally, continuum quantities that we compute with an input of experimental data, as described above, are indicated by a bar over the symbol, for example $\overline{\bf v}$.

\section {V.~Simplification of continuity equations for dust}

Here, we present our simplification of the continuity equations, Eqs.~(\ref{mass}-\ref{energy}), to describe our 2D dust layer. These simplifications involve three approximations suitable for the conditions in our 2D layer, and a treatment of two external forces, laser manipulation and gas friction, that are responsible for ${\bf f}_{\rm ext}$ and $P_{\rm ext}$. We describe these simplifications and the resulting continuity equations, next.

\subsection {A.~Approximations to Simplify Continuity Equations}

The first of our three approximations is $\partial / \partial t = 0$. This approximation is suitable for the steady overall conditions of our experiment. Aside from the particle-level fluctuations that one desires to average away, when adopting a continuum model, the only time-dependent processes in the experiment were the rastering of the laser beam at $> 100~{\rm Hz}$ and the $13.56~{\rm MHz}$ rf electric fields that powered the plasma. These frequencies are too high for the dust particles to respond, and the rastering of the beams is not a factor anyway because we will only use the continuum equations outside the laser beams.

The second approximation is that $\partial / \partial x$ is negligibly small. Due to the symmetry in our experiment design, as mentioned in Sec.~III, $x$ is treated as an ignorable coordinate, i.e., $\partial / \partial x = 0$. As a verification of this assumption, we observe that the ratio $(\partial \overline {v_x} / {\partial x}) / (\partial \overline {v_x} / \partial y)$ is of order $10^{-2}$, as a measure of the slightly imperfect symmetry of our experiment. Thus, for our experiment, we consider $\partial / \partial y$ to be of zeroth order and $\partial / \partial x$ to be two orders of magnitude smaller, when we approximate Eqs.~(\ref{mass}-\ref{energy}).

The third approximation is that $\overline {v_y}$ is negligibly small, based on our observation of the flow velocity of our 2D layer. Our results for the calculated flow velocity of the dust layer are shown in Fig.~2. From the velocity $\overline {v_x}$ in Fig~2(a), the flow can be easily identified from the two peaks with broad edges. However, the flow velocity in the $y$ direction is two orders of magnitude smaller than in the $x$ direction, as shown in Fig.~2(b).

\begin{figure}[htb]
\centering
\includegraphics{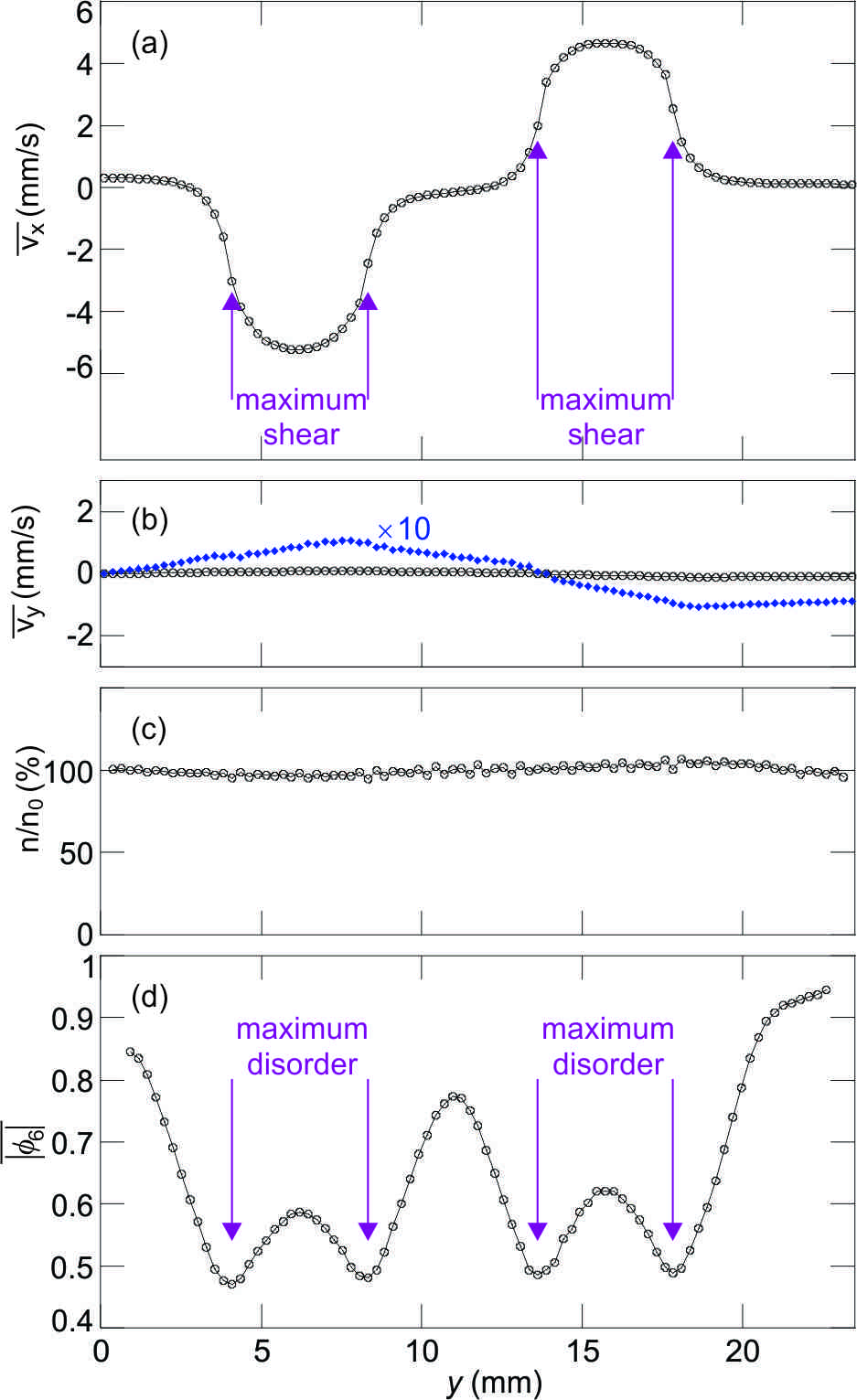}
\caption{\label{etasketch} (Color online). Profiles of continuum parameters during laser manipulation, including (a-b) flow velocity, (c) areal number density, and (d) a measure $|\phi_6|$ of local structural order that would be a value of 1 for a perfect crystal~\cite{Feng:10}. Disorder, as indicated by a small value in (d), is found to be greatest where the shear is largest, not where the flow is fastest.}
\end{figure}

Using the second and third approximations, and omitting terms that are small by at least two orders of magnitude, we can easily see that we can approximate ${\bf v} \cdot \nabla = 0$ and $\nabla \cdot {\bf v} = 0$. The latter indicates that the dust layer can be treated as an incompressible fluid in our experiment. Using these three approximations in Eq.~(\ref{mass}), we find that $\nabla \rho = 0$. In other words, for our approximations, the density $\rho$ is uniform, which is confirmed by our experimental observation of Fig.~2(c). Because the density is so uniform, we can also assume that plasma parameters, such as the dust charge $Q$, are also spatially uniform within the analyzed region.

In addition to these approximations, we also assume that $\eta$ and $\kappa$ are valid transport coefficients for our system. There are theoretical reasons to question whether 2D systems ever have valid transport coefficients, and this is typically tested {\it theoretically} using long-time tails in correlation functions~\cite{Ott:08, Donko:09}. It could be tested {\it experimentally} by repeating the determination of $\eta$ and $\kappa$ for vastly different length scales for the gradients of velocity and temperature and verifying that they do not depend on the length scale. However, such a test is not practical for experiments like ours, which tend to have a limited range of diameters of dust layers that can be prepared.

\subsection {B.~External Forces}

Now we consider external forces that contribute momentum and energy to the two equations, Eqs.~(\ref{momentum}) and (\ref{energy}). For our experiment, we can mention six {\it external} forces acting on the dust layer: gas friction, laser manipulation, electric confining force, gravity, electric levitating force, and ion drag. Since we only study the 2D motion of dust particles within their plane, the last three forces are of no interest because they are in the perpendicular direction, and will not affect the horizontal motion of particles that is of interest here. The electric confining force is balanced by the pressure inside the 2D dusty plasma lattice, $\rho_c {\bf E_{\rm conf}} = \nabla p$, as described in Sec.~II. Thus, only two of the six forces need to be considered: gas friction and laser manipulation.

Gas friction is the main dissipation mechanism in our experiment. We can consider the effect of this friction first at the level of a single dust particle, and then at the level of a continuum. For the {\it momentum} equation, we note first that at the particle level, a single dust particle moving at a speed of ${\bf v}_d$ experiences a drag acceleration of $- \nu_{\rm gas} {\bf v}_d$. At the continuum level, the contribution of this drag to Eq.~(\ref{momentum}) is simply the average acceleration experienced by all dust particles in a given spatial region,  $- \nu_{\rm gas} \overline{\bf v}$. For the {\it energy} equation, at the particle level the rate of energy dissipation for one dust particle is $- 2 \nu_{\rm gas} {\rm KE}_d$, which is the product of a drag force and velocity, where ${\rm KE}_d \equiv m_d {\bf v}_d^2/2$ is the kinetic energy of one dust particle. At the continuum level, averaging over all the dust particles in the given spatial region, the rate of energy loss per unit mass in Eq.~(\ref{energy}) is $- 2 \nu_{\rm gas} \overline{\rm KE} / m_d$~\cite{heating}.

\subsection {C.~Simplified Continuity Equations}

Using the three approximations listed above and taking into account gas friction and laser manipulation forces, the mass, momentum, and energy continuity equations become
\begin{equation}\label{mass_dp}
{\nabla \rho = 0}
\end{equation}
\begin{equation}\label{momentum_dp}
{\frac {\eta}{\rho} \frac{\partial^2 \overline{v_x}}{\partial y^2} -  \nu_{\rm gas} \overline{v_x} + f_{x\,{\rm laser}}= 0}
\end{equation}
\begin{equation}\label{energy_dp}
{\Phi + \frac {\kappa}{\rho} \frac{\partial^2}{\partial y^2}T - 2 \nu_{\rm gas} \overline{\rm KE} / m_d + P_{\rm laser} = 0,}
\end{equation}
where
\begin{equation}\label{viscous_heat}
{\Phi = \frac {\eta}{\rho} \left(\frac{\partial \overline{v_x}}{\partial y}\right)^2}
\end{equation}
is the viscous heating term, after simplifications based on the assumptions of $\partial / \partial x = 0$, $\overline {v_y} = 0$, and $\nabla \cdot {\bf v} = 0$.

In this paper, we will restrict our analysis to a spatial region where the laser force and power are zero, i.e., ${\bf f}_{\rm laser} = 0$ and $P_{\rm laser} = 0$. In this region, the momentum and energy continuity equations will be further simplified:
\begin{equation}\label{momentum_dpg}
{\frac{\partial^2 \overline{v_x}}{\partial y^2} - \frac {\rho \nu_{\rm gas}}{\eta}\overline{v_x}= 0}
\end{equation}
\begin{equation}\label{energy_dpg}
{\Phi + \frac {\kappa}{\rho} \frac{\partial^2}{\partial y^2}T - 2 \nu_{\rm gas} \overline{\rm KE} / m_d = 0.}
\end{equation}
To simplify the problem, we will assume that $\eta$ and $\kappa$ are independent of temperature, as discussed in~\cite{Nosenko:04, Nosenko:08}.

We can comment on the meaning of these two equations.  Equation~(\ref{momentum_dpg}) indicates a balance of the sideways transfer of momentum due to two mechanisms: viscosity arising from interparticle electric forces and frictional loss of momentum due to collisions with gas atoms. Equation~(\ref{energy_dpg}) describes the energy transferred from the viscous heating $\Phi$ and thermal conduction (the second term) as being balanced by the energy dissipated due to friction as expressed in the last term of Eq.~(\ref{energy_dpg}). We will use Eqs.~(\ref{momentum_dpg}-\ref{energy_dpg}) only in spatial regions where ${\bf f}_{\rm laser} = 0$ and $P_{\rm laser} = 0$, i.e., outside the laser beam. In the next section, we will present our calculation of terms appearing in Eqs.~(\ref{momentum_dp}-\ref{energy_dpg}) using the dust particles' position and velocity data from our experiment. The terms of interest in these equations are $\partial \overline{v_x} / \partial y$, $\partial^2 \overline{v_x} / \partial y^2$, $\overline{\rm KE}$, and $\partial^2 \overline{({\bf v} - \overline{\bf v})^2}/ \partial y^2$.

\section {VI.~Profiles of Quantities in the Continuity Equations}

\begin{figure}[htb]
\centering
\includegraphics{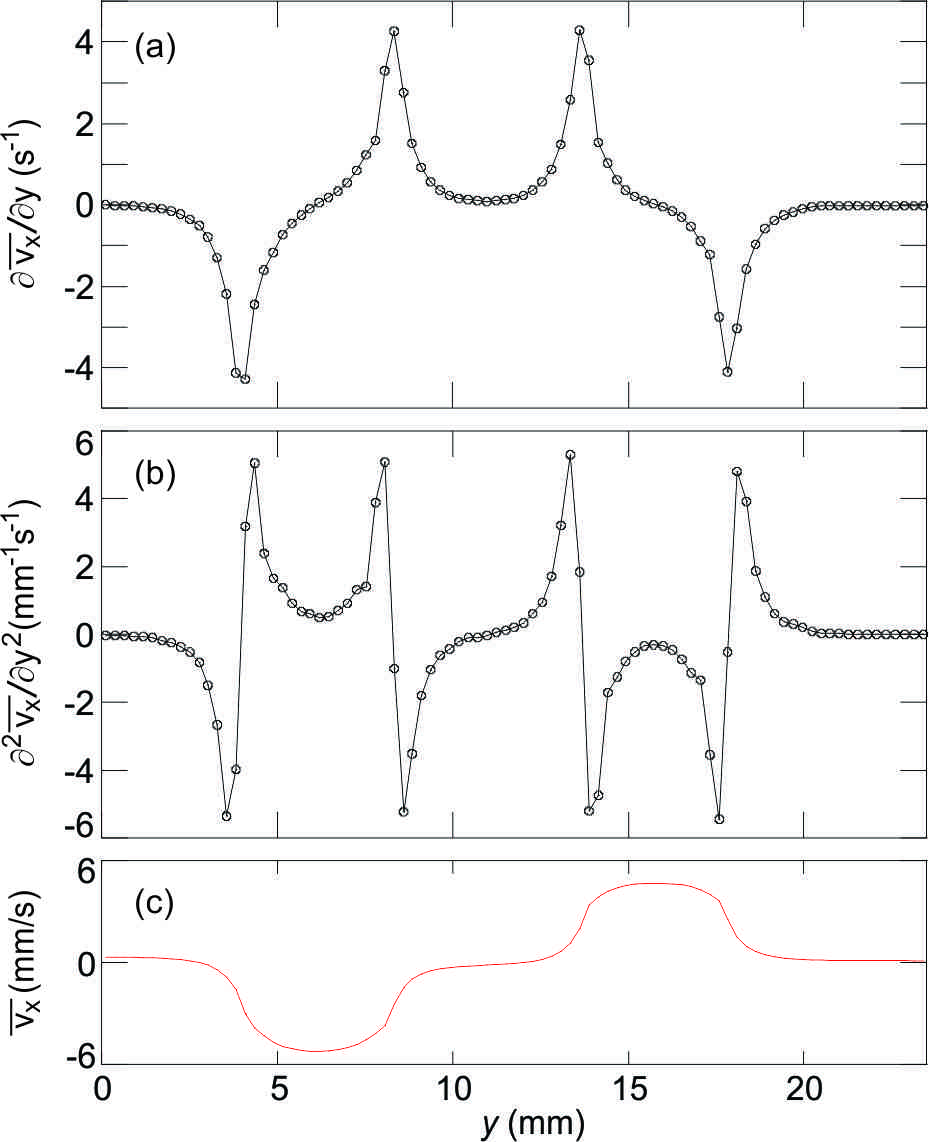}
\caption{\label{etasketch} (Color online). Profiles of the first (a) and second (b) derivatives of flow velocity. To make a comparison, the flow velocity profile $\overline{v_x}$ is also provided in (c).}
\end{figure}

Our results for the first and second derivatives of the flow velocity are presented in Fig.~3. These results are calculated using the flow velocity profile in Fig.~2(a), which we reproduce in Fig.~3(c). The first and second derivatives in Fig.~3(a-b) will be used in Eqs.~(\ref{viscous_heat}) and (\ref{momentum_dpg}), respectively. From Fig.~3(a), we can identify four points of maximum shear, i.e., maximum $\partial \overline{v_x} / \partial y$; these are at $y = 4.1~{\rm mm}$, $8.3~{\rm mm}$, $13.6~{\rm mm}$, and $17.8~{\rm mm}$. These points of maximum shear coincide with other features of interest: the minimum in the structural order, Fig.~2(d), and peaks in the mean-square velocity fluctuation profile, which we will present below. We find that disorder, as indicated by a small value in Fig.~2(d), is greatest where the shear is largest, not where the flow is fastest.

Comparing panels (b) and (c) of Fig.~3, we find that the profiles for the flow velocity $v_x$ and its second derivative are similar, in regions without laser manipulation, for example in the central region $8.3~{\rm mm} < y < 13.6~{\rm mm}$. This similarity is expected from the momentum equation, Eq.~(\ref{momentum_dpg}), provided that the viscosity $\eta$ is spatially uniform. We do not expect that $\eta$ would be spatially uniform since viscosity in general depends on temperature and the temperature is highly non-uniform, as we will show below. Nevertheless, we find that the two curves are nearly similar, with a small discrepancy that we will quantify below when we present data for the residual of Eq.~(\ref{momentum_dpg}).

\begin{figure}[htb]
\centering
\includegraphics{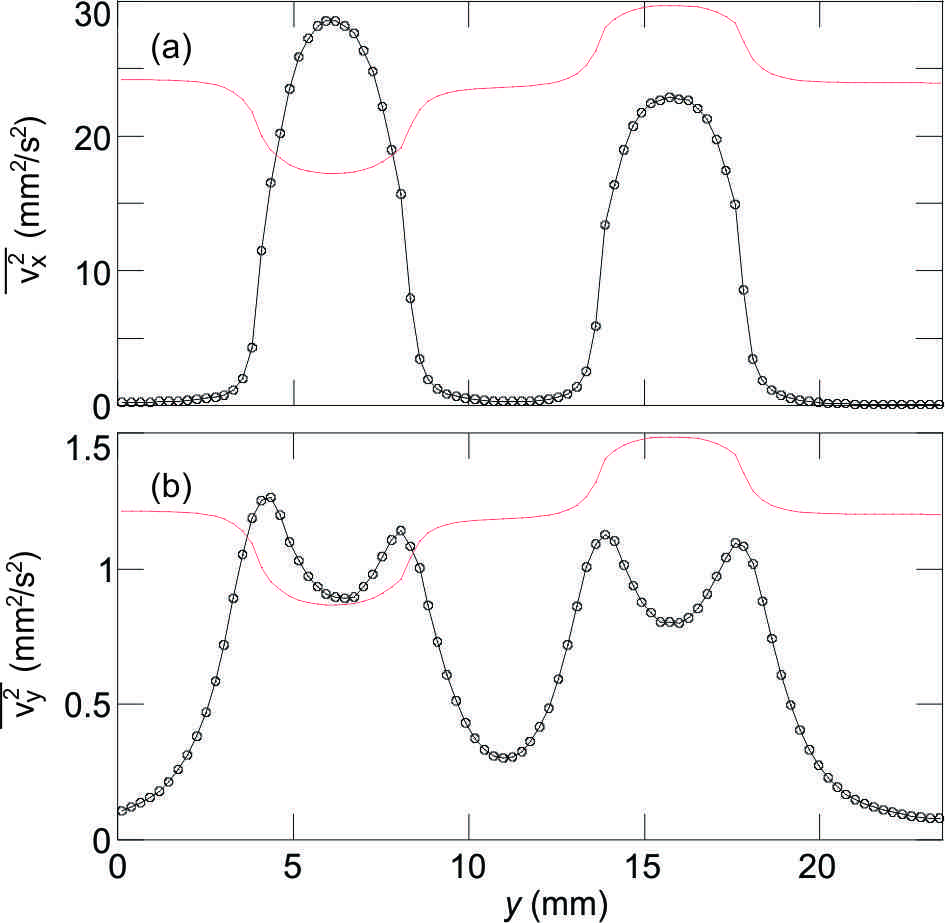}
\caption{\label{etasketch} (Color online). Profiles of the mean squared particle velocity shown separately for particle motion in the $x$ (a) and $y$ (b) directions. These quantities correspond to the mean particle kinetic energy $\overline{\rm KE}$, i.e., $(v_x^2+v_y^2)/2 = \overline{\rm KE}/m_d$. The thin curve in each panel is the flow velocity profile $\overline{v_x}$; its scale is shown in Fig.~3(c).}
\end{figure}

Profiles of the mean squared particle velocity, corresponding to the kinetic energy in the energy equation, Eq.~(\ref{energy_dpg}), are shown in Fig.~4. This kinetic energy includes energy associated with both the macroscopic flow and the fluctuations at the particle level. We will use these profiles in determining $\overline{\rm KE}$ in the next section, where we will find the residual of Eq.~(\ref{energy_dpg}).

\begin{figure}[htb]
\centering
\includegraphics{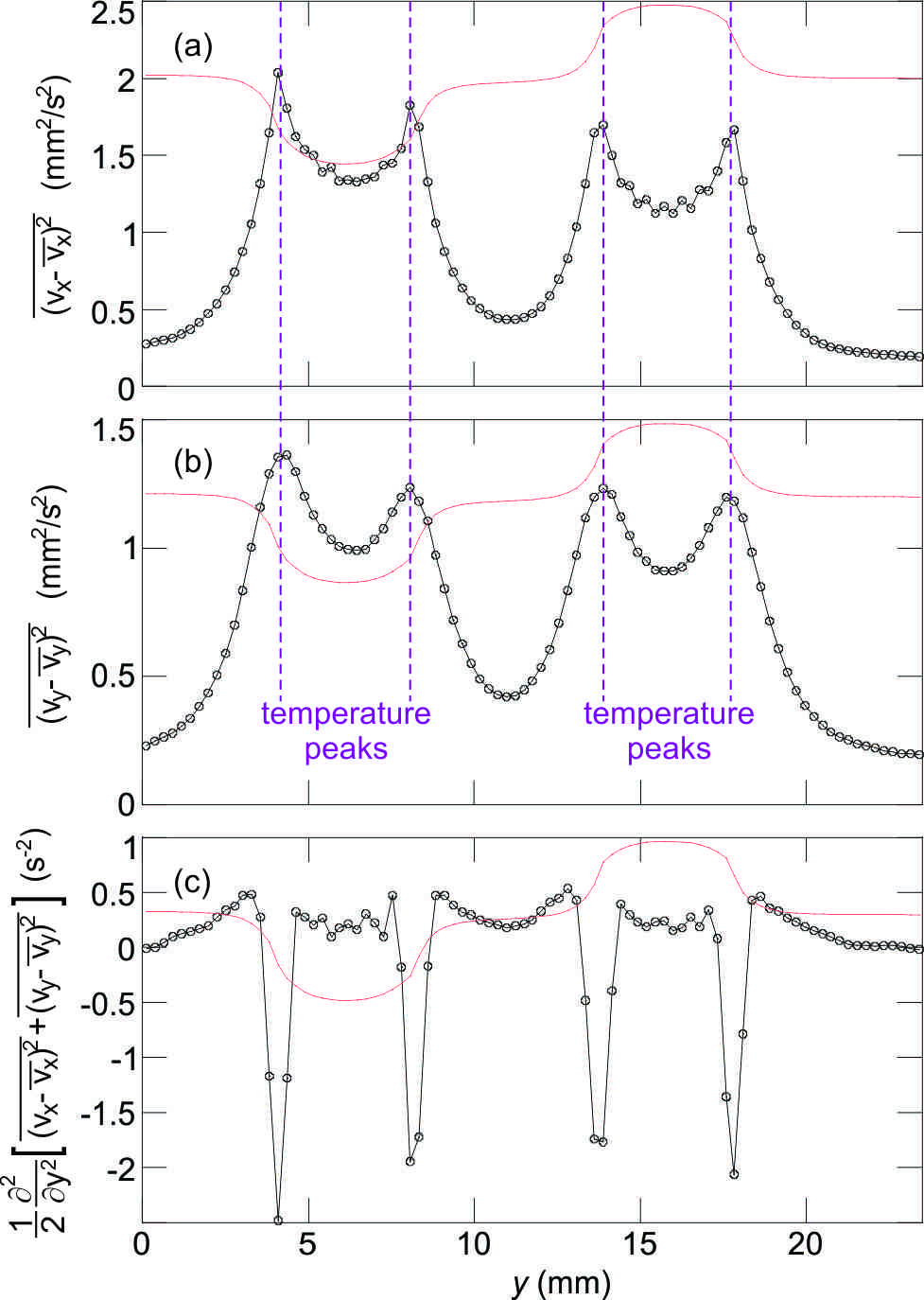}
\caption{\label{etasketch} (Color online). Profiles of the mean-square particle velocity fluctuation shown separately for particle motion in the $x$ and $y$ directions, (a) and (b), respectively. These quantities combined correspond to the kinetic temperature $T_{\rm kin}$, Eq.~(\ref{KT}). The profile of $T_{\rm kin}$ were reported in~\cite{Feng:12}, where we discovered peaks in the kinetic temperature where the shear is largest. The second derivative of the averaged mean-square particle velocity fluctuation for the motion in the $x$ and $y$ directions (c) will be used to determine the second derivative of the thermodynamic temperature in Eq.~(\ref{energy_dpg}). The thin curve in each panel is the flow velocity profile $\overline{v_x}$; its scale is shown in Fig.~3(c).}
\end{figure}

In Fig.~5, we present our results for the mean-square velocity fluctuation, which corresponds to the kinetic temperature as in Eq.~(\ref{KT}). We will use this kinetic temperature in place of the thermodynamic temperature $T$ in the energy equation, Eq.~(\ref{energy_dpg}). Unlike the kinetic energy $\overline{\rm KE}$, the kinetic temperature only includes the energy associated with the fluctuations of particle velocity about the flow velocity $\overline {\bf v}$~\cite{anisotropy}.

As reported in~\cite{Feng:12}, there are peaks in the kinetic temperature profile that coincide with the position of maximum shear. These peaks can also be seen in Fig.~5(a-b). In~\cite{Feng:12}, we attributed these peaks to viscous heating. As an intuitive explanation of viscous heating, consider that higher shear conditions lead to collisions of particles flowing at different speeds, causing scattering of momentum and energy that leads to higher random velocity fluctuations, and therefore higher kinetic temperature. In the next section, we provide further verification that the temperature peaks are due to viscous heating; we do this by confirming that three terms in the energy equation, including viscous heating, are in balance as indicated by their summing to zero.

\section {VII.~Results for Residuals of the Continuity Equations}

We now examine the momentum and energy equations, Eqs.~(\ref{momentum_dpg}) and (\ref{energy_dpg}), which are written so that the right-hand-side is zero. When we use these equations with an input of experimental data, however, the terms will not sum exactly to zero, but will instead sum to a finite residual. We calculate these residuals, and we vary two free parameters, the viscosity $\eta$ and the thermal conductivity $\kappa$, to minimize the residuals. (Specifically, we minimize the square residual summed over all bins in the central region of $8.6~{\rm mm} < y < 13.4~{\rm mm}$.) This minimization procedure yields the best estimation for the values for $\eta$ and $\kappa$, which will be our first chief result. We will then, as our second chief result, be able to make a spatially-resolved comparison of the magnitude of different terms in the energy equation, Eq.~(\ref{energy_dpg}).

\begin{figure}[htb]
\centering
\includegraphics{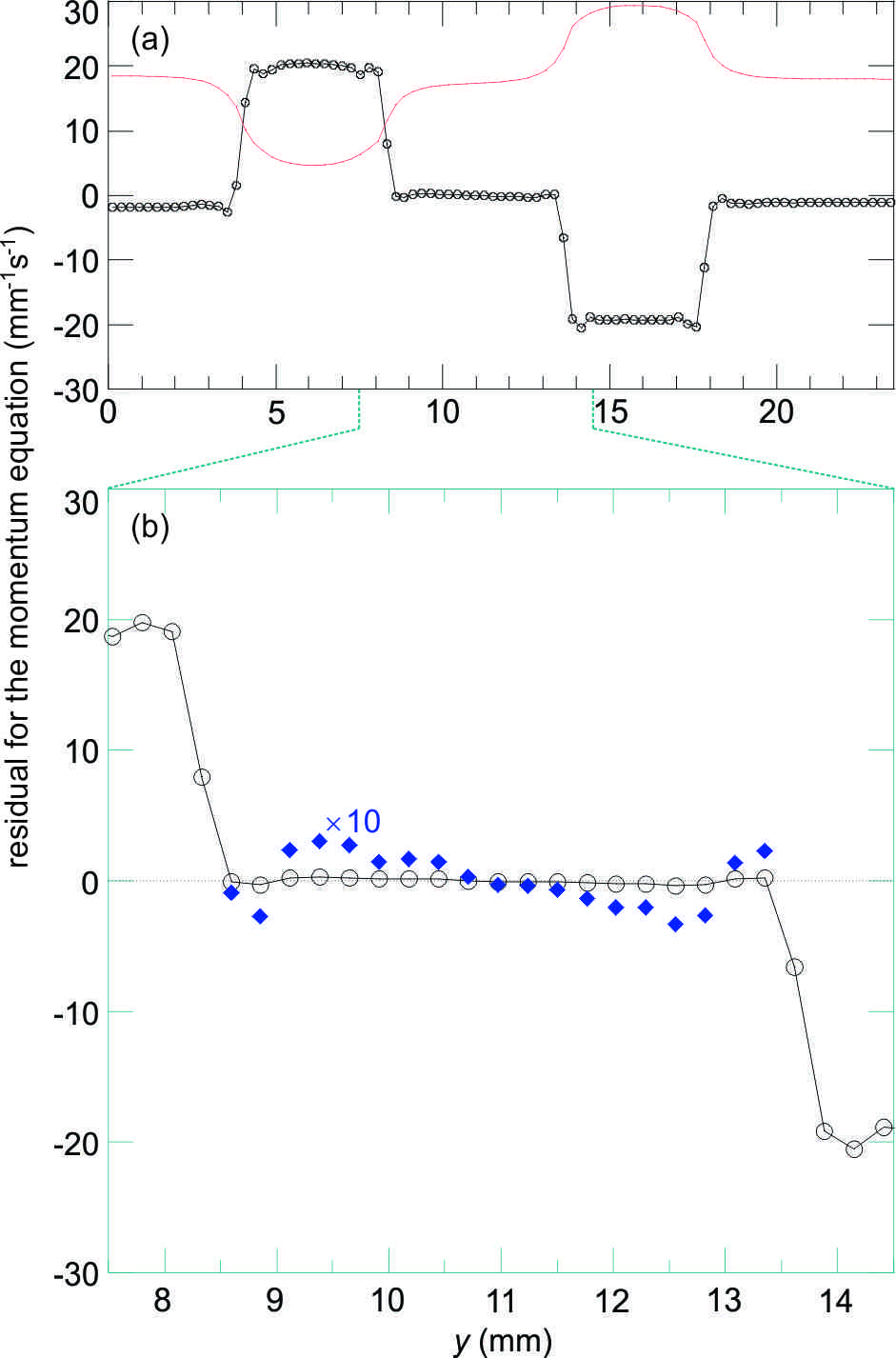}
\caption{\label{etasketch} (Color online). (a) A profile of the residual of the momentum equation, Eq.~(\ref{momentum_dpg}), assuming a kinematic viscosity of $\eta / \rho = 0.69~{\rm mm^2 / s}$. For the central region, magnified in (b), the summation of the squared residual reaches its minimum when $\eta / \rho = 0.69~{\rm mm^2 / s}$; this minimization process is how we determine $\eta$, which is one of our main results. The thin curve in (a) is the flow velocity profile $\overline{v_x}$; its scale is shown in Fig.~3(c).}
\end{figure}

Figure 6 shows the residual of the momentum equation, Eq.~(\ref{momentum_dpg}). This result is shown for $\eta/\rho = 0.69~{\rm mm^2/s}$, which is the best estimation of the kinematic viscosity. The data are shown as a spatial profile because we calculated Eq.~(\ref{momentum_dpg}) separately for each bin, i.e., each value of $y$. Two peaks in Fig.~6(a), located within the laser manipulation region, are due to the momentum contribution from the laser. We do not use Eq.~(\ref{momentum_dpg}) with these peaks because of a finite laser force ${\bf f}_{\rm laser}$ there. Instead we will use the flatter region between these peaks, where ${\bf f}_{\rm laser} = 0$, as magnified in Fig.~6(b). The small residuals in this flatter region indicate that the momentum equation, Eq.~(\ref{momentum_dpg}), is able to accurately account for the momentum of our 2D dust layer, and that the minimization process in this region yields a value for the viscosity.

\begin{figure}[htb]
\centering
\includegraphics{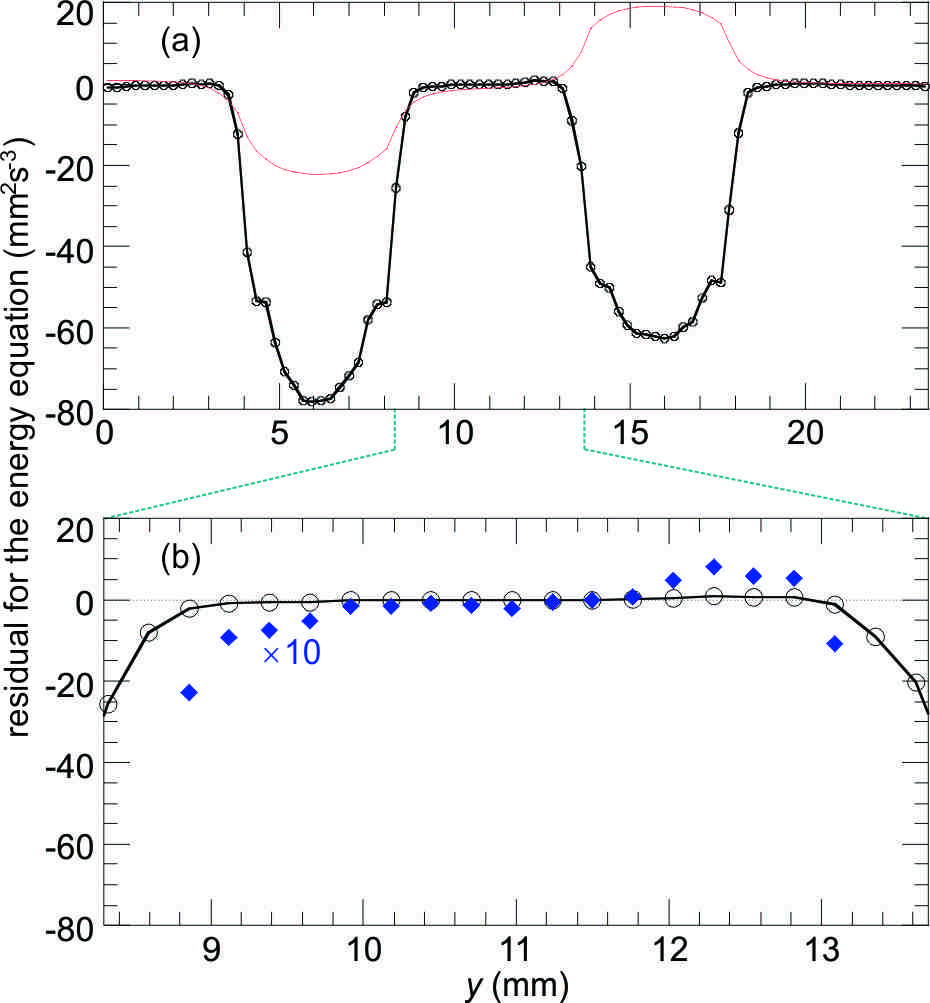}
\caption{\label{etasketch} (Color online). (a) A profile of the residual of the energy equation, Eq.~(\ref{energy_dpg}), assuming a thermal diffusivity of $\chi \equiv \kappa / c\rho = 8~{\rm mm^2 / s}$. For the central region, magnified in (b), the summation of the squared residual reaches its minimum when $\kappa / c\rho = 8~{\rm mm^2 / s}$. This minimization process is how we determine $\kappa$, which is another of our main results. The thin curve in (a) is the flow velocity profile $\overline{v_x}$; its scale is shown in Fig.~3(c).}
\end{figure}

Figure 7 shows the residual of the energy equation, Eq.~(\ref{energy_dpg}), for $\kappa/(c\rho) = 8~{\rm mm^2/s}$, which is the best estimation of the thermal diffusivity. In this calculation, we used the value of $\eta / \rho = 0.69~{\rm mm^2/s}$ from the momentum equation above, and we varied the value of $\kappa$ to minimize the residuals as described above. Two large negative peaks in Fig.~7(a) are due to the energy contribution from the laser manipulation $P_{\rm laser}$, which is not included in Eq.~(\ref{energy_dpg}). The small values of residuals in the flatter region between these peaks, as magnified in Fig.~7(b), show that the energy equation, Eq.~(\ref{energy_dpg}), accurately describe energy transport in our 2D dust layer, and that the minimization process in this region yields a value for the thermal diffusivity. By achieving a small value of the residual, we have verified that the three terms in the energy equation, Eq.~(\ref{energy_dpg}), are in balance. Since one of these terms is viscous heating and another is computed from the temperature profile which has peaks, the balance we observe here is consistent with the conclusion of~\cite{Feng:12} that the temperature peaks are due to viscous heating.

As the first chief result of this paper, we obtain the kinematic viscosity value of $\nu = \eta/\rho = 0.69~{\rm mm^2/s}$ and thermal diffusivity value of $\kappa/(c\rho) = 8~{\rm mm^2/s}$. These values are obtained simultaneously in a single experiment. A source of uncertainty in these values is systematic error in $\nu_{\rm gas}$, for example due to particle size dispersion or uncertainty in the Epstein drag coefficients~\cite{Liu:03}. This is so because our method actually yields results for $\eta / (\rho \nu_{\rm gas})$ and $\kappa / (c\rho \nu_{\rm gas})$. Another source of uncertainty is our simplification that we neglect heating sources other than laser manipulation~\cite{heating}; in a test we determined that $\kappa/(c\rho)$ is in a range from 7.5 to $8.4~{\rm mm^2/s}$, depending on whether these small heating effects are accounted for. We note that our results of $\eta / \rho$ and $\kappa / (c\rho)$ agree with the results of previous experiments~\cite{Nosenko:04, Nosenko:08} using the same size of dust particles and a similar value of $a/\lambda_D$.

\begin{figure}[htb]
\centering
\includegraphics{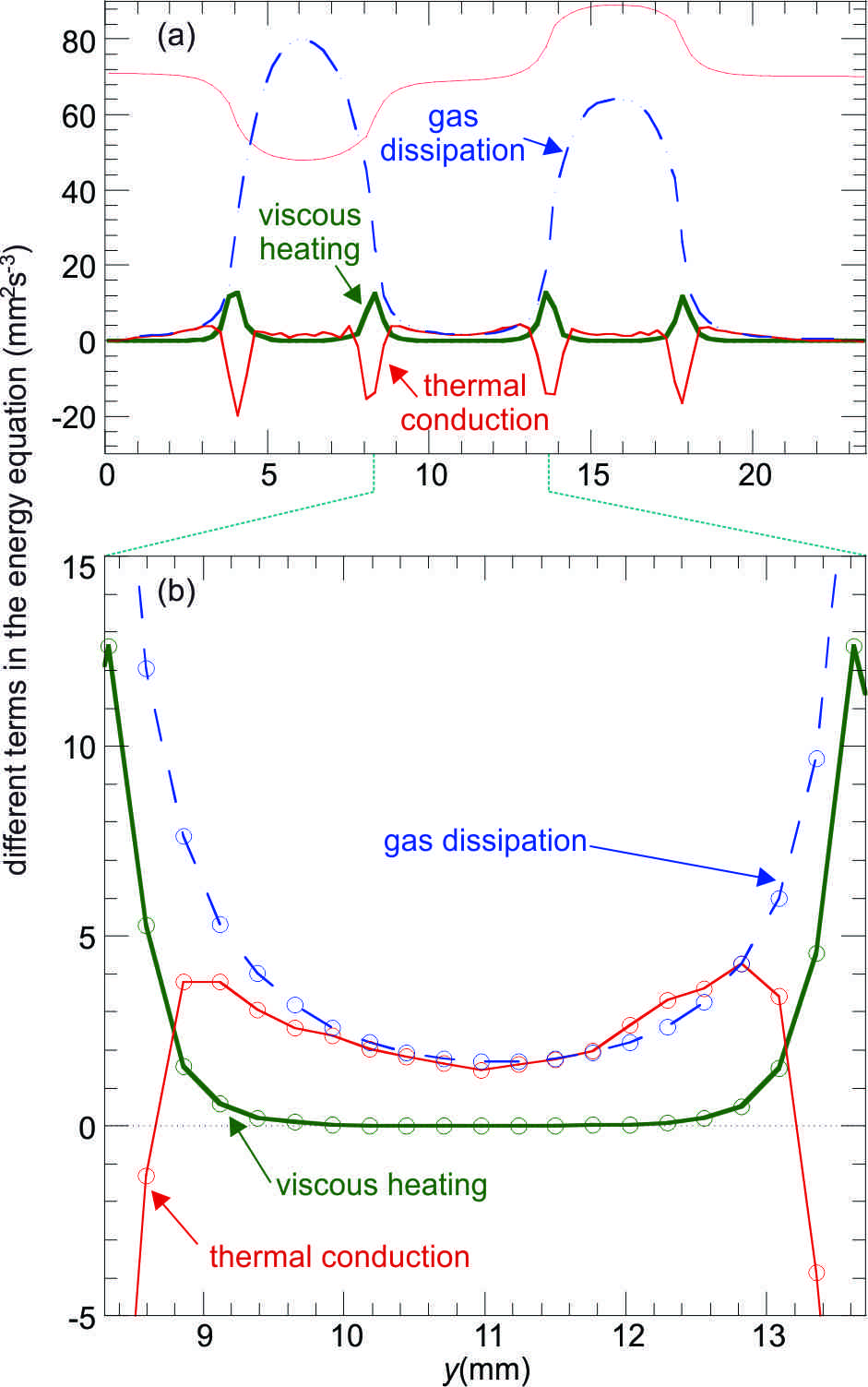}
\caption{\label{etasketch} (Color online). (a) Profiles of three terms in Eq.~(\ref{energy_dpg}), assuming the transport coefficients, $\eta$ and $\kappa$, obtained above. For the central region without laser manipulation (b), thermal conduction is one order magnitude larger than the viscous heating. This is our second chief result: a spatially-resolved comparison of the different mechanisms for energy transfer in our 2D dust layer. The thin curve in (a) is the flow velocity profile $\overline{v_x}$; its scale is shown in Fig.~3(c).}
\end{figure}

Our experiment allows us to obtain spatially-resolved quantitative measurements of three heat transfer effects: viscous heating, thermal conduction, and dissipation due to gas friction (i.e., cooling). These three effects appear as the three terms on the left-hand-side of Eq.~(\ref{energy_dpg}). As the second chief result of this paper, we plot these three terms, presented as spatial profiles, in Fig.~8.

Examining the spatial profiles for these terms, in Fig.~8, we see the most prominent features are two large peaks for the gas dissipation term where the flow velocity $\overline{v_x}$ is fastest; in this region the energy dissipation due to gas friction reaches its maximum. Despite the prominence of these features, however, they are not what interest us here. Instead, we are more interested in the regions of high shear, near the edge of the laser manipulation. Recall that in these high shear regions, the flow velocity gradient $\partial \overline{v_x} / \partial y$ is largest, and there are peaks in the profiles of the kinetic temperature and the second derivative of the flow velocity, Figs.~3 and 5.

In Fig.~8, our spatially-resolved profiles reveal that viscous heating and thermal conduction terms are peaked in regions of high shear. The viscous heating term, Eq.~(\ref{viscous_heat}), is always positive, meaning that viscous dissipation is always a source of heat wherever it occurs. The thermal conduction term partial $\partial^2 T/\partial y^2$, on the other hand, can be either positive or negative, indicating that heat is conducted toward or away from the point of interest, respectively. In locations where the shear is strongest, for example at $y = 4.1~{\rm mm}$, the thermal conduction term is negative indicating that heat is conducted away from that point.

Although viscous heating has great importance in all kinds of fluids, and it has been understood theoretically for a very long time~\cite{Brinkman:51}, a spatially-resolved measurement of it is uncommon. In most physical systems viscous heating is usually hard to measure either because the temperature increase is overwhelmingly suppressed by rapid thermal conduction, as we discussed in~\cite{Feng:12}, or because the thinness of the shear layer does not allow convenient {\it in-situ} temperature measurements. Most experimental observations of temperature increases due to viscous heating are either external or global measurements, and not spatially-resolved measurements like those that we report here. Indeed, in our literature search, we found no previous spatially-resolved experimental measurements of the viscous heating term, not only for dusty plasma, but also for any other physical system. As we explained in~\cite{Feng:12}, our ability to detect strong effects of viscous heating is due to the extreme properties of dusty plasma, as compared to other substances. Our ability to make spatially-resolved measurements is due to our use of video imaging of particle motion.

\begin{figure}[htb]
\centering
\includegraphics{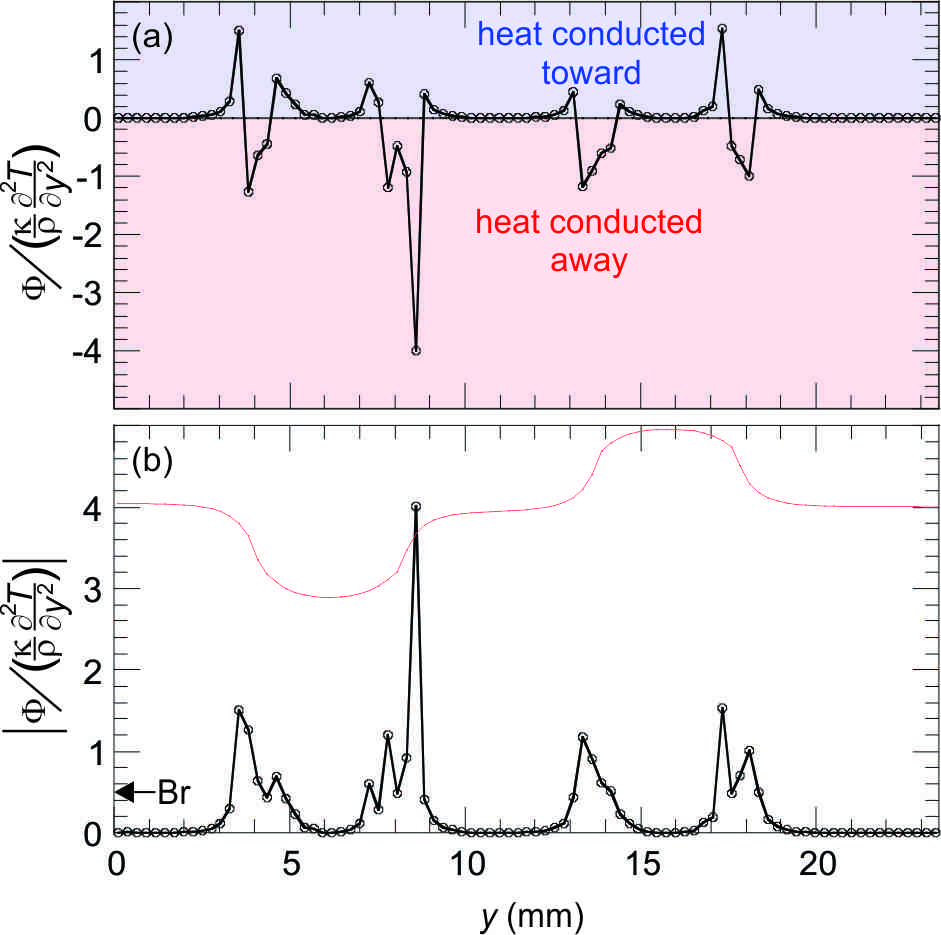}
\caption{\label{etasketch} (Color online). Profiles of (a) the ratio of the viscous heating and thermal conduction terms in Eq.~(\ref{energy_dpg}), and (b) the absolute value of this ratio. The sign of $\partial^2 T/\partial y^2$ determines the sign of the ratio in (a). A positive ratio indicates that heat is conducted toward the position of interest. The large values of the ratio in (b) at high shear regions indicate significant viscous heating at those locations. For comparison, in~\cite{Feng:12} we found a Brinkman number Br = 0.5 (indicated by the arrow), which is a global measure of the flow that provides less detailed information than the spatially resolved ratio shown here. The thin curve in (b) is the flow velocity profile $\overline{v_x}$; its scale is shown in Fig.~3(c).}
\end{figure}

To further analyze the second chief result of this paper, the spatial profiles of the viscous heating and thermal conduction terms in Fig.~8(a), we plot the ratio of these two terms in Fig.~9(a). This ratio has its largest positive and negative values in the regions of high shear. In Fig.~9(a), negative values of this ratio are observed to occur in the high shear regions, which indicates that heat is conducted away from these regions. This result is consistent with observation of kinetic temperature peaks here. To characterize the magnitude of these two terms, we plot in Fig.~9(b) the absolute value of this ratio. We can see that, within regions of high shear, this ratio can be as large as unity, or even larger. A typical value of this ratio in the shear region is of order 0.5 for our experiment. This matches the value of the Brinkman number, Br =0.5 that we found in~\cite{Feng:12} for the same experiment. The Brinkman number is a global measure of the viscous heating, in competition with thermal conduction.

\section {VIII.~Summary}
In summary, we reported further details of the laser-driven flow experiment in a dusty plasma that was first reported in~\cite{Feng:12}. We simplified the momentum and energy continuity equations, exploiting the symmetry and steady conditions of the experiment. We developed a method to obtain transport coefficients by minimizing the residuals of continuity equations using the input of experimental data. As our first chief result, we use this method to simultaneously determine, from the same experiment, two transport coefficients: kinematic viscosity and thermal diffusivity, which are based on viscosity and thermal conductivity, respectively. As our second chief result, we obtained spatially-resolved measurements of various terms in the energy equation. We found that, in a laser-driven dusty plasma flow, viscous heating is significant in regions with high shear, which is consistent with the interpretation of~\cite{Feng:12} that the peaks in the temperature profile are due to viscous heating.

This work was supported by NSF and NASA.


\begin{thebibliography}{16}
\bibitem{heating}
The external forces can add or remove energy from the collection dust particles, as expressed by $P_{\rm ext}$, which is the final term of the energy equation, Eq.~(\ref{energy}). Besides heating from laser manipulation and cooling from the gas friction as discussed in the text, we can mention two other contributions to $P_{\rm ext}$ for our experiment: heating from random kicks from collisions with the gas atoms and heating by fluctuating electric fields in the plasma. In the absence of any laser manipulation, the other three terms have their background levels which are in balance, so that their total $P_{\rm ext}$ is zero. In the presence of laser manipulation, even outside the region where the laser beams strike the dust particles, there are flows that result in a cooling by gas friction that is enhanced above its background level. When using the energy equation with data from our experiment with laser manipulation, we ignore two heating effects, gas atom collisions and fluctuating electric fields. Using data from an experimental run without laser manipulation, we estimated those effects, so that we could report the range of values for $\kappa/(c\rho)$ in the text.
\bibitem{anisotropy}
Our laser manipulation method introduces anisotropy in the particle velocities. Since our laser manipulation drives a flow only in the $\pm x$ directions, $|\overline{v_x}/\overline{v_y}| \gg 1$. Similarly, the kinetic energy is mostly due to the flow motion in the $x$ direction, i.e., $|\overline{v_x^2}/\overline{v_y^2}| \gg 1$, as shown in Fig.~4(a-b). However, the kinetic temperature due to the velocity fluctuations in the $x$ direction is almost the same as for fluctuations in the $y$ direction.
\end{thebibliography}
\end{document}